\documentclass{article}

\usepackage{PRIMEarxiv}

\usepackage[utf8]{inputenc} 
\usepackage[T1]{fontenc}    
\usepackage{hyperref}       
\usepackage{url}            
\usepackage{booktabs}       
\usepackage{amsfonts}       
\usepackage{nicefrac}       
\usepackage{microtype}      
\usepackage{lipsum}
\usepackage{fancyhdr}       
\usepackage{graphicx}       
\graphicspath{{media/}}     

\title{Generative Language Models on Nucleotide Sequences of Human Genes\thanks{This paper was originally published in Scientific Reports, Volume 14, Article Number 22204 (2024).}}

\author{
  Musa Nuri İhtiyar, Arzucan Özgür \\
  Department of Computer Engineering \\
  Boğaziçi University \\
  Istanbul\\
  \texttt{\{musa.ihtiyar, arzucan.ozgur\}@boun.edu.tr} \\
}

\begin{document}
\maketitle

\begin{abstract}
Language models, especially transformer-based ones, have achieved colossal success in natural language processing. To be precise, studies like BERT for natural language understanding and works like GPT-3 for natural language generation are very important. If we consider DNA sequences as a text written with an alphabet of four letters representing the nucleotides, they are similar in structure to natural languages. This similarity has led to the development of discriminative language models such as DNABert in the field of DNA-related bioinformatics. To our knowledge, however, the generative side of the coin is still largely unexplored. Therefore, we have focused on the development of an autoregressive generative language model such as GPT-3 for DNA sequences. Since working with whole DNA sequences is challenging without extensive computational resources, we decided to conduct our study on a smaller scale and focus on nucleotide sequences of human genes, i.e. unique parts of DNA with specific functions, rather than the whole DNA. This decision has not significantly changed the structure of the problem, as both DNA and genes can be considered as 1D sequences consisting of four different nucleotides without losing much information and without oversimplification. First of all, we systematically studied an almost entirely unexplored problem and observed that recurrent neural networks (RNNs) perform best, while simple techniques such as N-grams are also promising. Another beneficial point was learning how to work with generative models on languages we do not understand, unlike natural languages. The importance of using real-world tasks beyond classical metrics such as perplexity was noted. In addition, we examined whether the data-hungry nature of these models can be altered by selecting a language with minimal vocabulary size, four due to four different types of nucleotides. The reason for reviewing this was that choosing such a language might make the problem easier. However, in this study, we found that this did not change the amount of data required very much.
\end{abstract}


\section{Introduction}\label{sec1}
DNA is a significant building block of living organisms, responsible for carrying fundamental data about various factors affecting the life of the living beings to which it belongs, such as humans, animals, plants and others. DNA consists of nucleotides, yet some subsequences, called genes, have unique functionalities and special meaning.

Genes can be represented in textual format as sequences of four different types of nucleotides, namely adenine, thymine, cytosine and guanine, as shown in Figure~\ref{fig:symbolicgenes}. The sequence length can vary from 100 to 10,000. Consequently, despite looking simple, the number of possible nucleotide sequences for a chosen size like 10,000 is vast (about $10^{6000}$). Yet, only a few exist in humans, animals and other living beings.  If we focus specifically on the genes that occur in humans, the number drops to about 70,000.

\begingroup
\begin{figure}[ht]
\vskip\baselineskip 
	\centering
		\includegraphics[width=6cm, keepaspectratio]{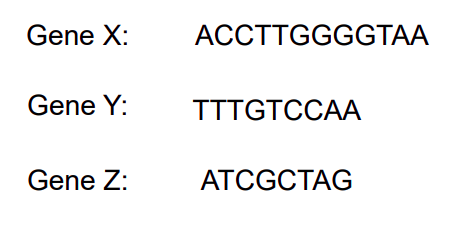}
		\caption{Nucleotide Sequences For Symbolic Genes.}
		\label{fig:symbolicgenes}
\end{figure}
\endgroup

Artificial intelligence and machine learning are among the most popular and exciting fields of our time. These fields involve helping computers or robots perform operations usually done by human beings, such as recognizing objects in an image, understanding natural language in the form of text, playing games like chess, or driving a car. However, they can also be used for more sophisticated goals, like helping humans make new scientific discoveries.

Because DNA nucleotide sequences resemble the text format of natural languages, methods commonly used for natural language processing could theoretically be used to solve various genomics tasks. Although the use of natural language understanding techniques, which involve understanding a particular sentence, is an option, the use of natural language generation methods, which is about modeling language using given sentences, is also possible in the field of genomics.

In this study, we investigated the problem of modeling nucleotide sequences for genes associated with humans. In other words, we looked at the performance of different machine learning models in the task of looking at nucleotide sequences for some of the genes to obtain a model so that they can predict how likely it is that unseen sequences belong to a human gene. They could also generate new nucleotide sequences for candidate genes. The expected behavior is illustrated in Figure~\ref{fig:modelpipeline}.

\begingroup
\begin{figure}[ht]
\vskip\baselineskip 
	\centering
		\includegraphics[width=12cm, keepaspectratio]{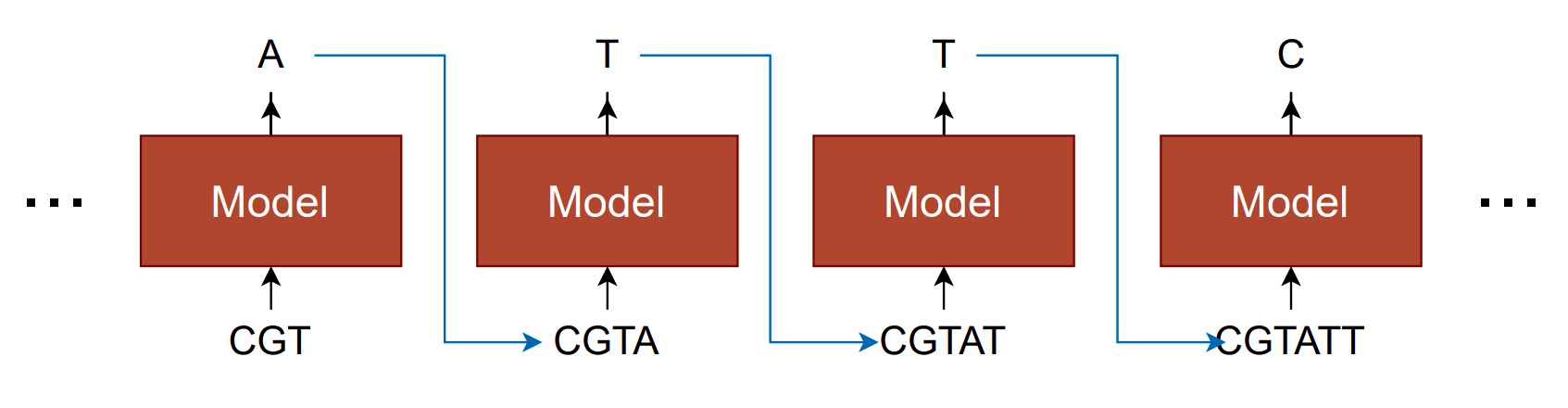}
		\caption{An Autoregressive Generative Language Model For Genes.}
		\label{fig:modelpipeline}
\end{figure}
\endgroup

Generative models have proven their success in different domains, most obviously with ChatGPT \cite{ChatGPT}. Theoretically, they can carry out almost all of the tasks done by a discriminative model. Especially combining different domains such as natural language and programming yielded incredible outcomes like OpenAI Codex \cite{chen2021evaluating}, which can write code by only getting natural language as input. DALL-E 2 \cite{ramesh2022hierarchical} is another nice example from a different field. One of its most exciting achievements is replacing an object in an image with another thing without changing the background, called in-painting. Similar work can be done for the combination of any domain and natural language. Although the current version of our study only focuses on genes, the same idea could be applied to whole DNA sequences, and a similar combination with natural language could provide huge benefits. To give concrete examples, we might show a mutated DNA sequence causing disease and prompt it to convert this DNA sequence into the healthy version. Another option could be giving a sequence carrying a specific property like being resistant as input and wanting the program to describe how to integrate that feature into another organism's DNA so that it has the same property as well. In fact, theoretical limits for what can be done may be very close to the limits of what genomics could do at all. When all these possibilities are taken into account, we had better think carefully about ethical considerations too.

In Section 2, we discuss some related studies. That is, we give some concrete and specific examples on related areas to better understand the current research direction. Then, in Section 3, we describe how the dataset we worked on was obtained and give various statistics about the dataset. In Section 4, we explain in detail how the experiments were conducted and the results obtained. Finally, conclusions and future work are discussed in Sections 5 and 6, respectively.

\section{Related Work}\label{sec2}

The invention of computers hasn’t affected only well-known areas such as digital sectors. It was very crucial for other fields like science in general as well. Due to the power of fast and reliable calculations they provided, computers were beneficial for lots of different scientific and technological processes.

Biology benefits from these machines with great potential. Various biological building blocks could be studied much more robustly and quickly with these solid computing machines \cite{yang2020review, li2021bioseq}. Such efforts are sometimes referred to as bioinformatics or computational biology.

Proteins are one of the most fundamental building blocks when it comes to biology. Both discriminative and generative models are used for these critical elements. While McDowall and Hunter \cite{mcdowall2011interpro} provide a tool for automatic classification of proteins using models such as hidden Markov models, Madani et al. \cite{madani2020progen} attempt to develop a generative model for proteins based on big data that can be used for various application domains like medicine.

Since proteins can have different 3D structures, working with them requires special care. In addition, there are studies that focus on the interactions between proteins and various chemicals, as these are of great importance for topics such as drug development. For example, Nagamine and Sakakibara \cite{nagamine2007statistical} use Support Vector Machines, a conventional machine learning algorithm, to predict the probability of interaction for a given protein and chemical pair. On the other hand, Cobanoglu et al. \cite{cobanoglu2013predicting} use probabilistic matrix factorization, which is an alternative method for a closely related task.

DNA and RNA sequences have a massive importance in our lives; consequently, applications related to them are prevalent. The area is not new, even though various facilities in the past were limited. For example, Wang et al. \cite{wang1999new} devised two algorithms, one unsupervised, for classifying DNA. Nonetheless, Nguyen et al. \cite{nguyen2016dna} prefer using convolutional neural networks for classifying DNA even though convolutional neural networks are not commonly used for classifying sequence-like inputs. Furthermore, Qi et al. \cite{qi2020clustering} use an unsupervised clustering method on RNA sequences in addition to supervised applications.

Some studies used classical artificial intelligence methods to solve bioinformatics problems without using machine learning. Altschul et al. \cite{altschul1990basic} use such an approach to measure the similarity of two sequences, as this is a crucial problem for tasks like search. Burge and Karlin \cite{burge1997prediction} use a probabilistic approach to solve tasks such as gene identification. Lowe and Eddy \cite{lowe1997trnascan} use the same technique for RNA-related work with a very low error rate. Also, Trapnell et al. \cite{trapnell2013differential} develop a robust method for gene expression by taking advantage of RNA sequences. Despite the existence of such studies, the integration of machine learning techniques in one form or another is extremely popular nowadays.

Traditional machine learning methods are a serious option for tasks with limited data, as deep learning approaches, their main competitors, need more data to perform better. Rannala and Yang \cite{rannala2003bayes} pursue a statistical method for estimating different species using DNA. Choi et al. \cite{choi2017evaluation} investigate the success of the classical logistic regression algorithm in RNA sequence studies.

As in other fields, deep learning is inevitably very common in bioinformatics because it is very successful when enough data is available for a given task. Lan et al. \cite{lan2018survey} mention several use cases for this, which is expected since deep learning is extremely popular in this field. As an interesting example, Zeng et al. \cite{zeng2023evaluating} deep learning to make predictions about RNA and protein pairs to better understand these structures.

There are some other works using machine learning on proteins. AIPs-SnTCN \cite{raza2023aips} uses a combination of embedding techniques to make classification on peptides. Moreover, cACP-DeepGram \cite{akbar2022cacp} uses combinations of deep learning methods for a cancer-related study.

Lastly, using language models is also a popular topic. For instance, DNABert \cite{ji2021dnabert} uses an approach similar to the classical BERT, where data comes from human DNA for training and various DNA sources for testing instead of natural language in order to solve various tasks related to genomics. While DNABert worked on datasets where samples were DNA sequences, our study was on a smaller scale since our samples were nucleotide sequences of genes. 

In addition, Luo et al. \cite{luo2022biogpt} try to do a similar study for the generative part of the coin, yet for biomedical text instead of directly using biological units. However, a successful study which trained a gigantic generative transformer model on structures like DNA or proteins is not available, to the best of our knowledge. 

Like natural language processing, biomedical domains try to use language models as much as possible due to their vast potential and proven success in several applications. This circumstance is also called biological language models. It is possible to find very different use cases, ranging from discriminative to generative when the subject is examined in depth \cite{osmanbeyoglu2011n, wang2019high, li2021bioseq, benegas2022dna, gankin2023species, liang2012segmenting}.

As a result, bioinformatics is a very hot field, and our study focuses directly on DNA instead of other molecules such as RNA or proteins or interactions between pairs. In addition, currently, popular machine learning-based approaches, especially deep learning, are employed in our work. Also, in spite of the fact that DNABert is one of the closely related studies to our interests, the task is selected as a generative problem, not a discriminative one, since the latter was studied more extensively and the former has, arguably, more importance. Lastly, our work is relatively small sized in terms of datasets and models owing to factors like limited resources. However, it is still work on an exciting and unexplored problem. 

To provide a summary, Transformer model \cite{vaswani2017attention} provided huge improvements in Natural Language Processing. This is valid for both discriminative models like BERT \cite{devlin2018bert}, RoBERTa \cite{liu2019roberta} and generative models like GPT \cite{radford2018improving}, GPT-2 \cite{radford2019language}, GPT-3 \cite{brown2020language}. Owing to similarity between natural languages and biological languages, they were tried on biological data like DNA and Protein sequences as well, yet it was done mainly for discriminative cases. DNABERT \cite{ji2021dnabert}, ProteinBERT \cite{brandes2022proteinbert} and DNABERT-2 \cite{zhou2023dnabert} gave successful results in these domains. These crucial milestones from Natural Language Processing or Bioinformatics are listed chronologically in Table \ref{table:importantworks}.

\begin{table}[]
\centering
\caption{Critical Works Related To Our Paper}
\vskip\baselineskip
\begin{tabular}{|c|c|}
\hline
\textbf{Model} & \textbf{Year} \\ \hline
Transformer \cite{vaswani2017attention}    & 2017          \\ \hline
GPT \cite{radford2018improving}  & 2018          \\ \hline
BERT \cite{devlin2018bert}          & 2018          \\ \hline
GPT-2 \cite{radford2019language} & 2019 \\ \hline
RoBERTa  \cite{liu2019roberta}      & 2019          \\ \hline
GPT-3 \cite{brown2020language} & 2020 \\ \hline
DNABERT  \cite{ji2021dnabert}      & 2021          \\ \hline
ProteinBERT \cite{brandes2022proteinbert}   & 2022          \\ \hline
DNABERT-2 \cite{zhou2023dnabert} & 2023 \\ \hline
\end{tabular}
\label{table:importantworks}
\end{table}

\section{Dataset}\label{sec3}
The aim of preparing the dataset was to use the nucleotide sequences for genes existing in human beings. The National Center for Biotechnology Information (NCBI) \cite{ncbi}, a government institution, provides different databases such as Genome, Gene, and Nucleotide. From the Gene Database, we filtered genes such that only the ones existing in human beings remained. Then we got information like the commonly used name, id used in the database, and nucleotide sequence for all these genes. Nucleotide sequences for a specific gene can be accessed using the Nucleotide database of NCBI, and the dataset we obtained is online \cite{datasets}. Some general statistics about the dataset are given in Table \ref{table:datasetstatistics} to provide a general overview of what kind of data we worked on.

\begin{table}[]
\centering
\caption{General statistics about the dataset}
\vskip\baselineskip

\begin{tabular}{|c|c|c|c|}
\hline
                           & \textbf{Whole Dataset} \\ \hline
\textbf{Number of Samples} & 70475                  \\ \hline
\textbf{Min Length}         & 2                      \\ \hline
\textbf{Max Length}         & 2473603                \\ \hline
\textbf{Mean Length}        & 26417.25               \\ \hline
\textbf{Median Length}      & 3566.0                 \\ \hline
\end{tabular}
\label{table:datasetstatistics}
\end{table}

Next, we had to filter out genes whose sequence length exceeded a certain threshold, since transformer-based models assume a certain maximum length. This threshold was set at 1000, since larger values led to problems in fitting the models to the memory of the computers. Then we used 80\% of it for training and 20\% of the dataset for testing purposes. The split was done randomly. Some general statistics on the datasets can be found in Table \ref{table:filtereddatasetstatistics}. The datasets are available online \cite{datasets}.

\begin{table}[]
\centering
\caption{General statistics about the train and test datasets}
\vskip\baselineskip
\begin{tabular}{|c|c|c|c|}
\hline
                           & \textbf{Actual Dataset} & \textbf{Training/Validation Set} & \textbf{Test Set} \\ \hline
\textbf{Number of Samples} & 22886                  & 18308                 & 4578            \\ \hline
\textbf{Min Length}         & 2                      & 2                     & 4                \\ \hline
\textbf{Max Length}         & 1000                   & 1000                  & 1000           \\ \hline
\textbf{Mean Length}        & 365.51                 & 364.35                & 370.13          \\ \hline
\textbf{Median Length}      & 295.0                  & 295.0                 & 295.0            \\ \hline
\end{tabular}
\label{table:filtereddatasetstatistics}
\end{table}

In addition to these operations, the use of a validation set for models requiring hyperparameter optimization was necessary. For this reason, the training set was further split into 75\% and 25\%, where the former corresponds to the training set without validation and the latter is the validation set. 

The nucleotide sequence was the essential information for each gene, yet other details such as NCBI gene id, symbol, description, and gene type information were also given. Lastly, $<$ and $>$ symbols were used to represent the beginning and end of the genes on the nucleotide sequence column of the corresponding files, and processing was done accordingly in the experiments.

\section{Experiments and Results}\label{sec4}
The dataset and the code used for implementing the ideas described in the paper is reproducible and available online at \url{https://github.com/boun-tabi/GenerativeLM-Genes} for all different techniques used.

\subsection{N-gram Language Models}

For N-gram language models \cite{shannon1948mathematical}, we experimented with six different values of N, starting from one and going up to six. The perplexity values obtained are given in Table \ref{table:ngramresults}. Moreover, Natural Language Toolkit (NLTK) library \cite{nltk} was helpful for both using N-grams and calculating perplexity.

\begin{table}[]
\centering
\caption{Results for different n values}
\vskip\baselineskip
\begin{tabular}{|c|c|c|c|}
\hline
\textbf{N Value} & \textbf{Perplexity} \\ \hline
1                & 3.9959              \\ \hline
2                & 3.9214              \\ \hline
3                & 3.8985              \\ \hline
4                & 3.8834              \\ \hline
5                & 3.8723              \\ \hline
6                & $\infty$            \\ \hline
\end{tabular}
\label{table:ngramresults}
\end{table}

\subsubsection{N-gram Language Models with Laplace Smoothing}

Getting the perplexity of infinity when N is selected as six was an interesting result since the training set contained thousands of genes having thousands of nucleotides, so encountering an unseen sequence with length six did not look normal. Nevertheless, when the cause of the situation is inspected, it is noticed that the problem was related to understanding the beginning or end of the gene, which is a crucial component for modeling.

Then Laplace smoothing \cite{lidstone1920note} is used to avoid this problem. Different N values are tried starting from six and going up to eight. The corresponding perplexity values obtained are given in Table \ref{table:laplaceresults}.

\begin{table}[h]
\centering
\caption{Laplace smoothing results for different n values}
\vskip\baselineskip
\begin{tabular}{|c|c|c|c|}
\hline
\textbf{N Value} & \textbf{Perplexity} \\ \hline
6                & 3.8616              \\ \hline
7                & 3.8390              \\ \hline
8                & 3.8263              \\ \hline
\end{tabular}
\label{table:laplaceresults}
\end{table}

The best model we obtained had about 66,000 parameters.

\subsection{Long Short-Term Memory Based Language Model}

After experimenting with N-gram language models, we continued with deep learning-based approaches. The first option was using Long Short-Term Memory \cite{hochreiter1997long} based models. For getting input, a layer for learning embedding and position information was used at the beginning. Then a single or multiple Long Short-Term Memory layers are used depending on hyperparameter configuration. Lastly, the aim was predicting the next word; because of this, a linear layer with the size of vocabulary many units are used with Softmax to get predictions. The model architecture is visualized in Figure~\ref{fig:rnnarchitecture}. Adam was chosen as the optimizer to use. 

\begingroup
\begin{figure}[ht]
\vskip\baselineskip 
	\centering
		\includegraphics[width=12cm, keepaspectratio]{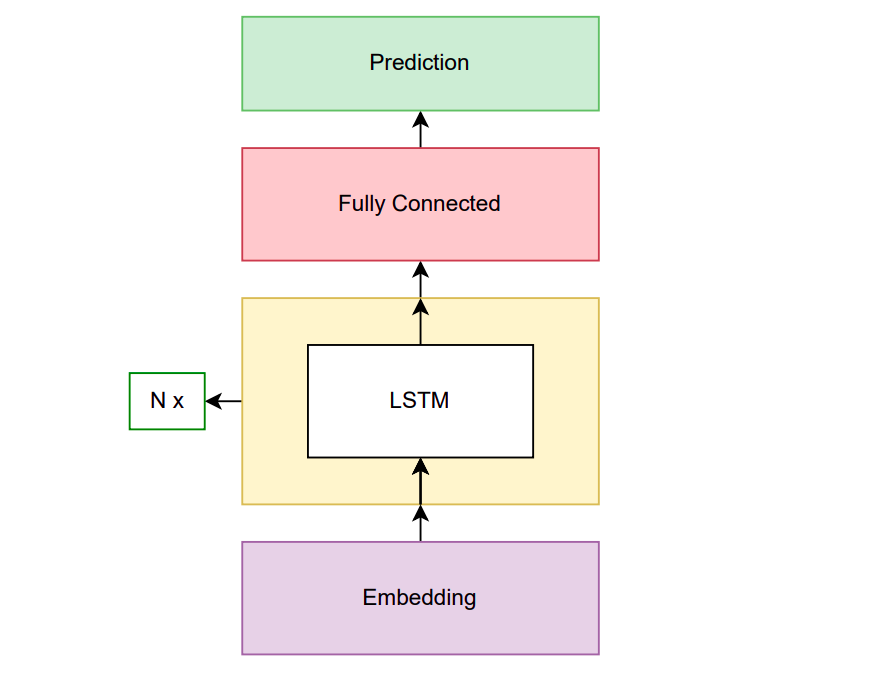}
		\caption{Long Short-Term Memory Based Architecture.}
		\label{fig:rnnarchitecture}
\end{figure}
\endgroup

Hyperparameter optimization was a crucial component. The basic procedure followed is like the following. We started with a hyperparameter configuration which looked reasonable. After that, if the model was underfitting, relevant hyperparameters are changed so that the model's capacity is increased. When overfitting was seen, the opposite is done. The details can be seen in Table S1. 

There were three main hyperparameters to play with. Embedding dimension, the number of LSTM layers to use and LSTM dimension. For any of these hyperparameter values, the model was run for 40 epochs. Then the result where the absolute difference between training and validation perplexities was less than 0.02 was chosen as the result of that hyperparameter trial. The corresponding perplexity values obtained for the best model are given in Table \ref{table:rnnresults}.

\begin{table}[]
\centering
\caption{Long short-term memory result}
\vskip\baselineskip
\begin{tabular}{|c|c|c|c|}
\hline
\textbf{N Value} & \textbf{Validation Set Perplexity} & \textbf{Test Set Perplexity} \\ \hline
1024                & 3.2726 & 3.2878              \\ \hline
\end{tabular}
\label{table:rnnresults}
\end{table}

The best model we obtained had 8,140,552 parameters.

\subsection{Transformer Based Language Model}

After using Long Short-Term Memory based models, transformer-based models were the next candidate. Actually, most of the procedures were very similar to the previous case. For getting input, a layer for learning embedding and position information was used at the beginning. Then a single or multiple transformer blocks are used depending on the hyperparameter configuration. Lastly, the aim was predicting the next word; because of this, a linear layer with the vocabulary size many units are used with Softmax to get predictions. The model architecture is visualized in Figure~\ref{fig:transformerarchitecture}. Adam was chosen as the optimizer to use. 

\begingroup
\begin{figure}[ht]
\vskip\baselineskip 
	\centering
		\includegraphics[width=12cm, keepaspectratio]{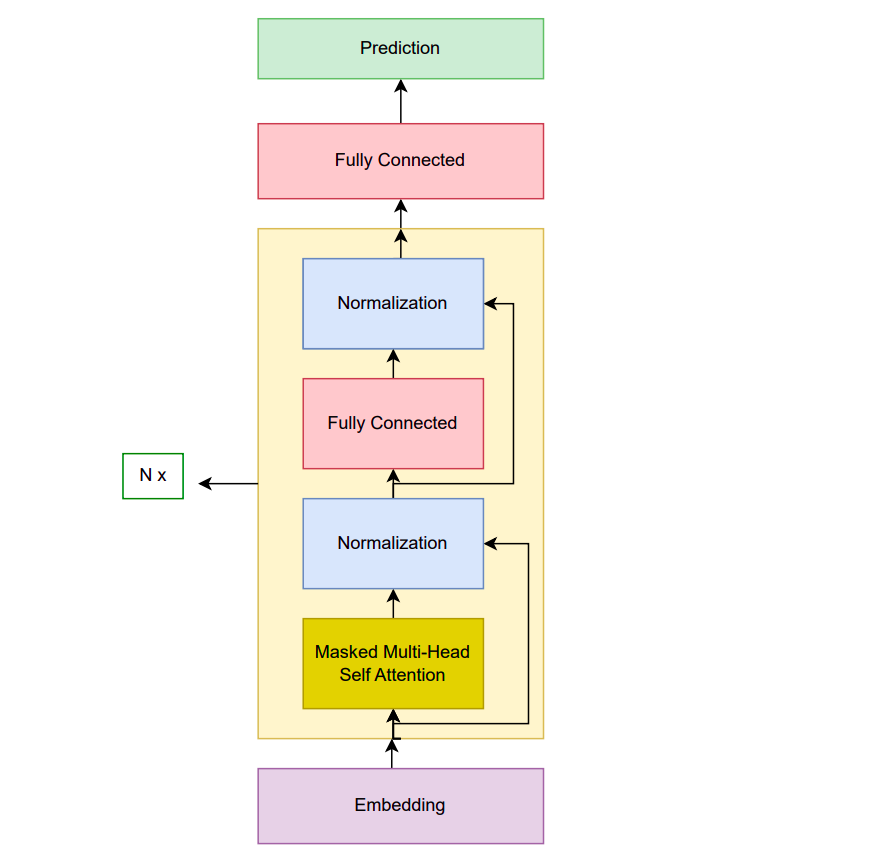}
		\caption{Transformer Architecture.}
		\label{fig:transformerarchitecture}
\end{figure}
\endgroup

Hyperparameter optimization was a crucial component. The basic procedure followed is like the following. We started with a hyperparameter configuration which looked reasonable. After that, if the model was underfitting, relevant hyperparameters are changed so that the model's capacity is increased. When overfitting was seen, the opposite is done. The details can be seen in Table S2. 

There were four main hyperparameters to play with. Embedding dimension, the number of transformer blocks to use transformer feed forward dimension and number of heads to use for a certain transformer block. The model was run for 40 epochs for any of these hyperparameter values. Then the result where the absolute difference between training and validation perplexities was less than 0.02 was chosen as the result of that hyperparameter trial. The corresponding perplexity values obtained for the best model are given in Table \ref{table:transformerresults}.

Lastly, Tensorflow library \cite{tensorflow}, its high-level API Keras \cite{keras} and its NLP tool KerasNLP \cite{kerasnlp} were very beneficial for implementing stuff related to Deep Learning. The best model we obtained had 1,844,988 parameters. 

\begin{table}[]
\centering
\caption{Transformer result}
\vskip\baselineskip
\begin{tabular}{|c|c|c|c|}
\hline
\textbf{N Value} & \textbf{Validation Set Perplexity} & \textbf{Test Set Perplexity} \\ \hline
1024                & 3.6880 & 3.6905              \\ \hline
\end{tabular}
\label{table:transformerresults}
\end{table}

The results obtained for different models based on test set perplexity is visualized in Figure~\ref{fig:testperp} using Matplotlib library \cite{Hunter:2007}.

\begingroup
\begin{figure}[ht]
\vskip\baselineskip 
	\centering
		\includegraphics[width=12cm, keepaspectratio]{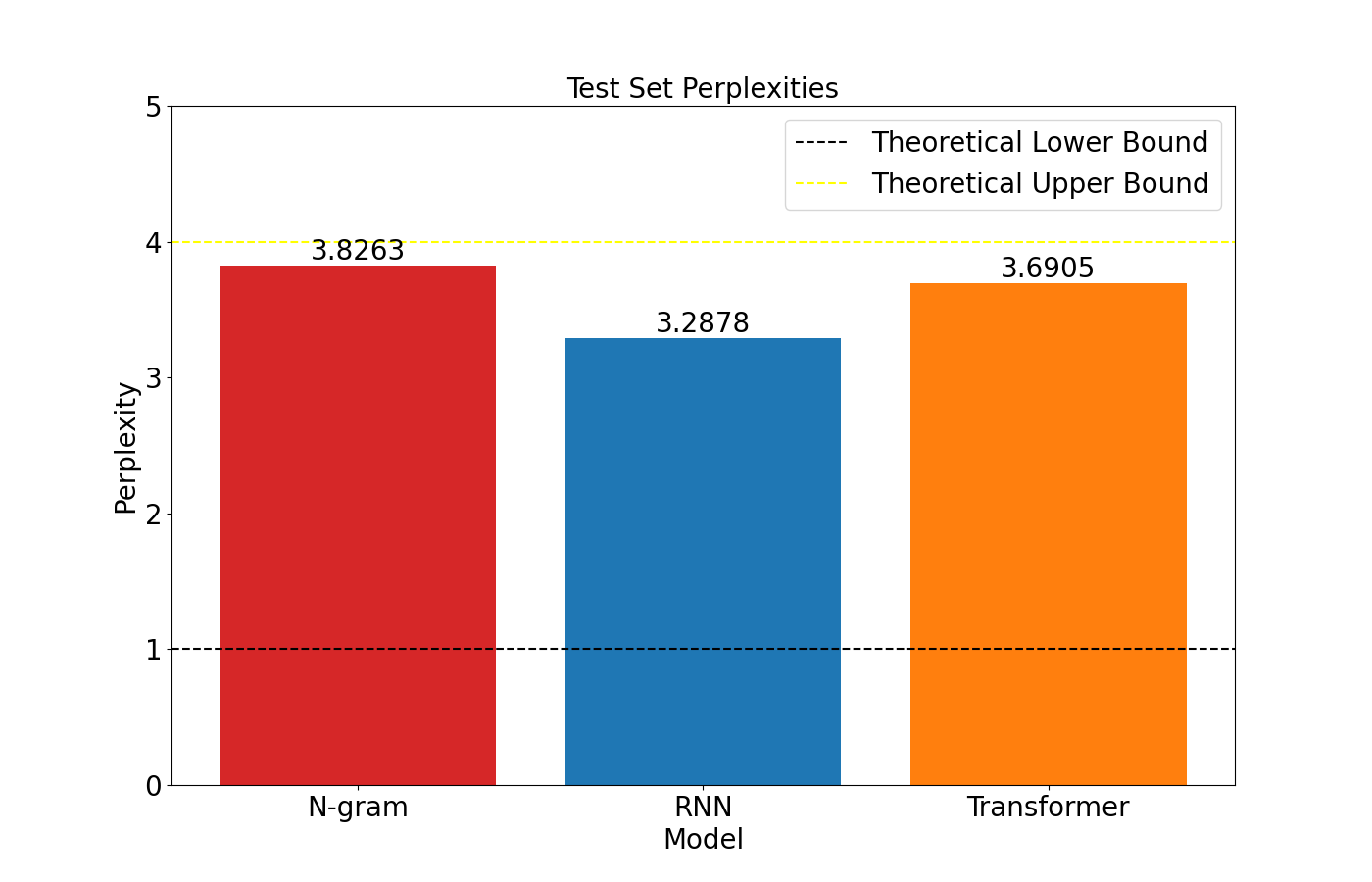}
		\caption{Comparison of Different Methods based on Test Set Perplexity.}
		\label{fig:testperp}
\end{figure}
\endgroup

\subsection{Performances On Real Life Tasks}

Evaluating the success of the methods is not trivial, so a real-life task could be very useful. For this reason, we tried to find a real-life problem where modeling nucleotide sequences of human genes could be helpful. Distinguishing between real nucleotide sequences of a real gene and a mutated version seemed like an exciting task, since the ability to do such a task well requires some level of understanding of the data we were working on. In fact, what we did is similar to the evaluation of GPT-2 \cite{radford2019language} on tasks like Winograd Schema Challenge \cite{levesque2012winograd}, in the sense that our model is trained as a generative model, but it could be adopted to do discriminative tasks when the correct setting is prepared.

Then the strategy was to obtain pairs such that each pair consists of the nucleotide sequence of an actual human gene and its mutated version. Then obtained models would assign a probability or perplexity to both of them and predict which one is real and which one is mutated based on these values. If it uses probability, the one with a higher predicted probability would be selected as the real one. If the model uses perplexity, the gene with a lower value will be chosen as the real one since the lower, the better for the perplexity metric. Then the accuracy for predictions made will be checked to see the success of a certain model. After deciding on this strategy, the only part that needed to be obtained was obtaining a dataset of real and mutated gene pairs.

\subsubsection{Synthetic Mutation Dataset}

One way to obtain a dataset consisting of real and mutated pairs is basically generating the mutated samples from the normal ones with the help of a computer by using a random process.

The method followed was like the following. Firstly, some hyperparameters such as the number of samples to obtain for the dataset and the maximum number of changes to be applied to a selected real gene, are determined. We selected the first as 1000. For the second, we experimented with three different values, which are 1, 5, and 10 to see how the performance changes according to this parameter. Lastly, since the process contains random number generation, a random seed was used to obtain reproducible results. For that part, we, again, used three different seeds to be able to use the general behavior.

After determining hyperparameters, for each selected sample from the test set, the number of changes to apply was selected, which is between 1 and the maximum number of changes (both inclusive) with equal probability. Then for each change, an option among three different candidates was chosen with equal probability. These options were adding a new randomly chosen nucleotide to a random position, deleting an arbitrary position in the sequence, and changing the nucleotide on a randomly selected position to another one. The described operations are visualized in Figure~\ref{fig:mutationoperations} and the mentioned algorithm is also described in Figure~\ref{fig:alg}.

\begingroup
\begin{figure}[ht]
\vskip\baselineskip 
	\centering
		\includegraphics[width=12cm, keepaspectratio]{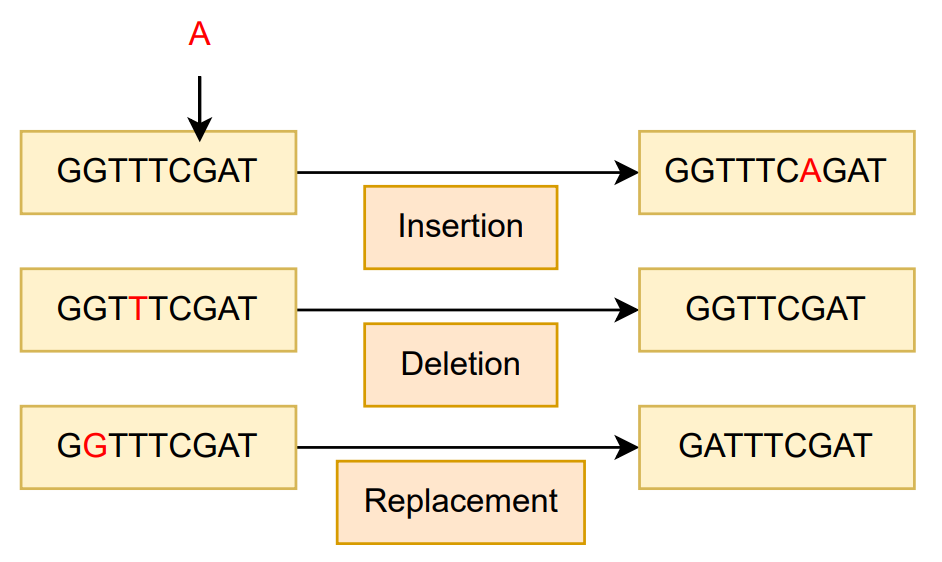}
		\caption{Synthetic Mutation Dataset Operation Types.}
		\label{fig:mutationoperations}
\end{figure}
\endgroup

\begingroup
\begin{figure}[ht]
\vskip\baselineskip 
	\centering
		\includegraphics[width=12cm, keepaspectratio]{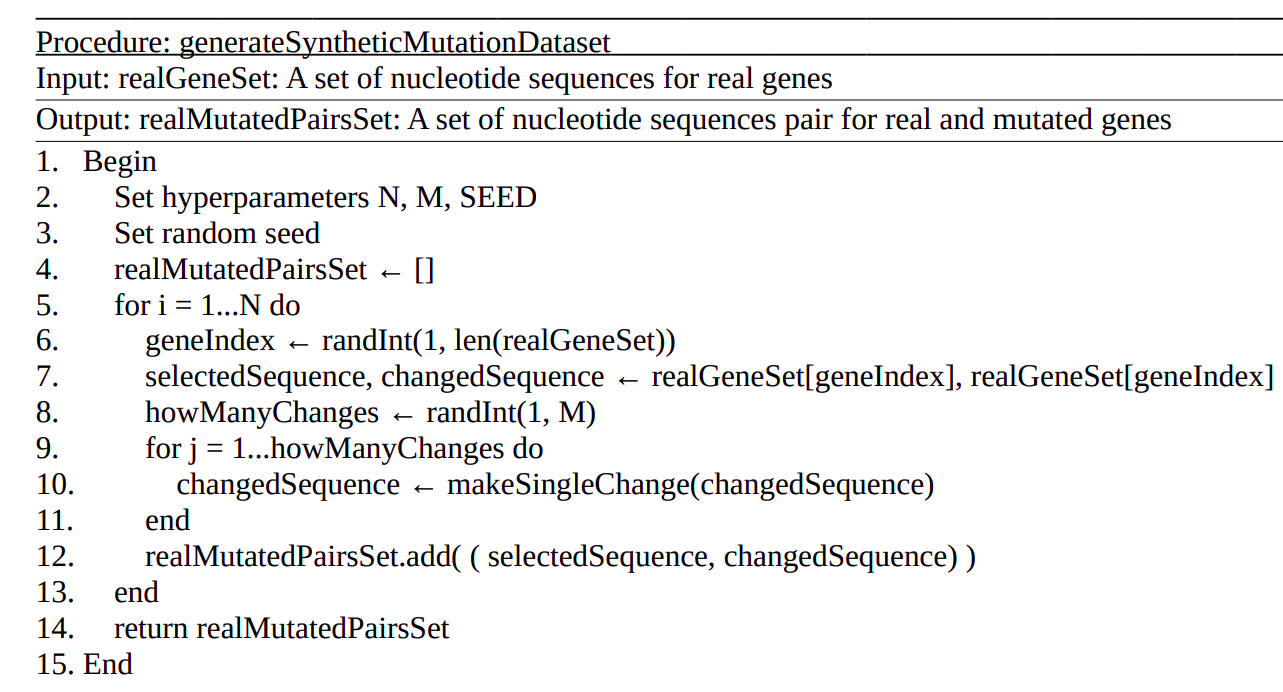}
		\caption{Procedure for Generating a Synthetic Mutation Dataset.}
		\label{fig:alg}
\end{figure}
\endgroup

As stated, this procedure was carried out for all three different values of the maximum number of mutations and three different seeds, leading to nine different synthetic mutation datasets. 
For synthetically obtained nine different datasets, containing real and mutated gene pairs, the best models from N-gram with Laplace smoothing, recurrent neural networks and transformer-based language models were chosen. Their predictions were taken for each couple and the accuracy was calculated for each of them. The results obtained based on perplexity for three different random seeds, which were 419432, 623598 and 638453 in the code are given in Table \ref{table:syntheticperplexityresults}.

\begin{table}[]
\centering
\caption{Synthetic mutation dataset result based on perplexity predictions}
\vskip\baselineskip
\begin{tabular}{|c|c|c|c|}
\hline
\textbf{Model} & \textbf{Max Change: 1} & \textbf{Max Change: 5} & \textbf{Max Change: 10}   \\ \hline
N-gram            & (0.649, 0.647, 0.653) & (0.749, 0.741, 0.729)  & (0.787, 0.793, 0.782)   \\ \hline
RNN               & (0.728, 0.745, 0.721) & (0.816, 0.785, 0.814)  & (0.840, 0.846, 0.853)   \\ \hline
Transformer       & (0.638, 0.652, 0.665) & (0.711, 0.700, 0.690)  & (0.766, 0.745, 0.753)   \\ \hline
\end{tabular}
\label{table:syntheticperplexityresults}
\end{table}

The results obtained for different models based on synthetic mutation dataset perplexity is visualized in Figure~\ref{fig:syntheticaccperp} using the Matplotlib library.

\begingroup
\begin{figure}[ht]
\vskip\baselineskip 
	\centering
		\includegraphics[width=12cm, keepaspectratio]{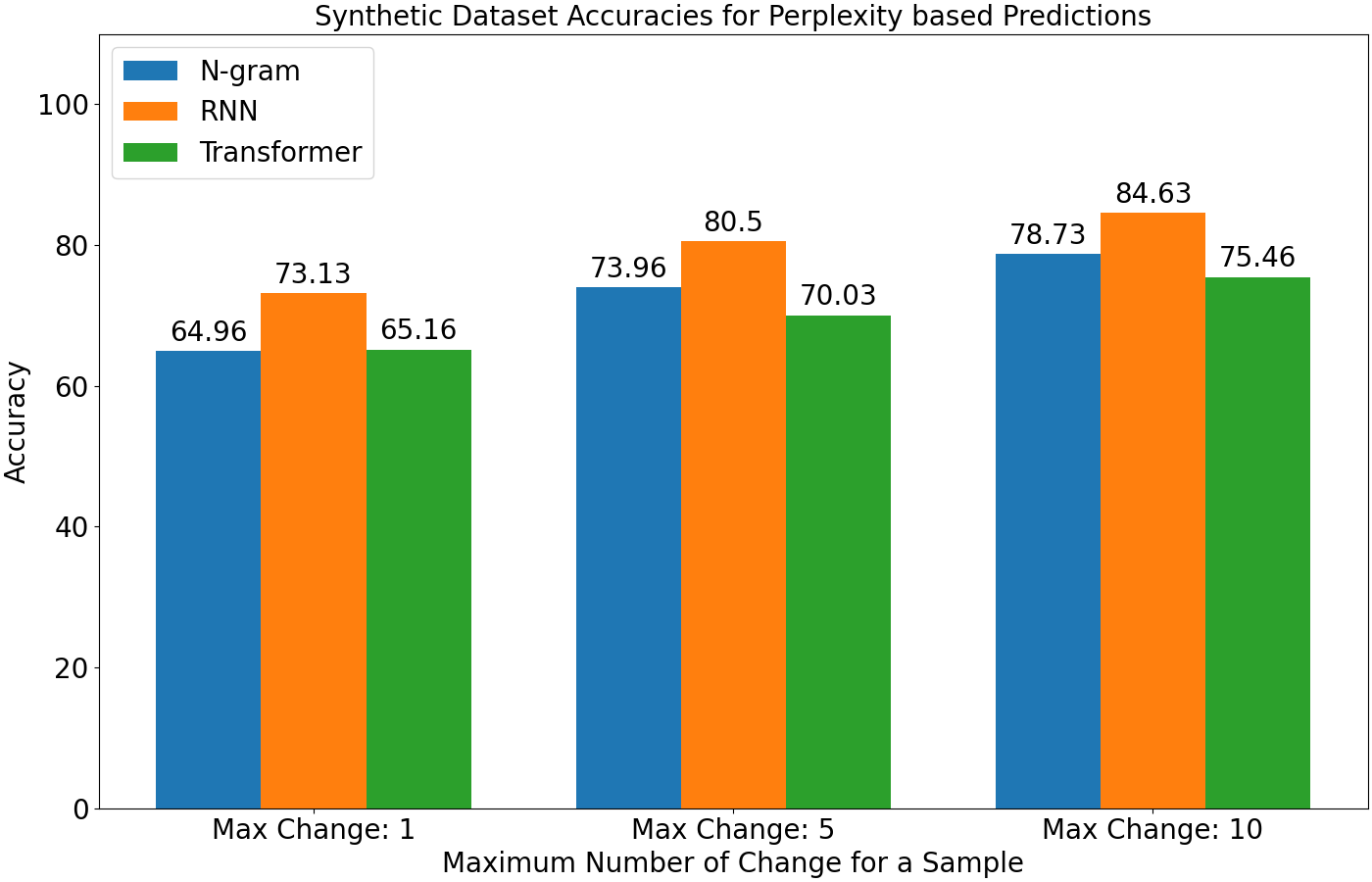}
		\caption{Averaged Accuracy Results Using Perplexity for Predictions.}
		\label{fig:syntheticaccperp}
\end{figure}
\endgroup

\subsubsection{Real Mutation Dataset}
Although synthetic datasets are beneficial for measuring the success of different alternatives, checking the performance on the real data is also very important. Consequently, finding mutations for genes in our test set was needed. 

Human Gene Mutation Database \cite{stenson2003human, stenson2014human, stenson2020human} was a nice resource with almost no alternative for our goal. Even though the number of mutations we could find for genes in our test set was only 10, which is very limited, it still helped to show that our models could make nice predictions for real examples as well. Since the database is not publicly available, we will, unfortunately, not be able to share the obtained mutation dataset.

Moreover, details about which model made a mistake on which example is shown in Table \ref{table:realmutationdetails}. More specific information about this part of the study, such as where the best model made a mistake, can be seen in Table \ref{table:realmutationdetails}.

\begin{table}[h]
\centering
\caption{Correctness of perplexity based predictions for different models}
\vskip\baselineskip
\begin{tabular}{|c|c|c|c|c|}
\hline
\textbf{Gene Number} & \textbf{Mutation Information} & \textbf{N-gram} & \textbf{RNN} & \textbf{Transformer} \\ \hline
\textbf{1}           & 1 Substitution                & False            & True         & False                \\ \hline
\textbf{2}           & 1 Substitution                & True             & True         & True                 \\ \hline
\textbf{3}           & 1 Insertion                   & True             & True         & False                \\ \hline
\textbf{4}           & 1 Insertion                   & False            & True        & False                 \\ \hline
\textbf{5}           & 1 Substitution                & True             & True         & True                 \\ \hline
\textbf{6}           & 1 Substitution                & True             & False         & True                 \\ \hline
\textbf{7}           & 1 Substitution                & False            & True         & False                \\ \hline
\textbf{8}           & 1 Substitution                & False            & False         & True                 \\ \hline
\textbf{9}           & 1 Substitution                & True             & True         & False                \\ \hline
\textbf{10}          & 10 Deletions                  & True             & True         & True                 \\ \hline
\end{tabular}
\label{table:realmutationdetails}
\end{table}

\section{Conclusion}\label{sec5}
In this study, we used different natural language processing-based language modeling techniques to model nucleotide sequences of human genes. For big datasets, expecting better performance from deep learning-based methods, primarily transformer-based language models were standard; however, the problem we focused on was inherently the opposite. In other words, the actual genes existing in nature are limited. However, the language in nucleotide sequences looked relatively simple because there were only four words excluding the beginning and end of genes. Consequently, it was interesting to examine whether or not this phenomenon will reduce the gap for deep learning-based models.

When the results obtained are examined, it is clear that simple models such as N-gram language models performed better than their transformer-based competitors and performed really close to recurrent neural network competitors. Also, it was seen that classical metrics used to measure success were not perfect in the sense that N-gram models performed better than transformer-based models in real-world tasks, although they looked worse when the perplexity metric was used for the comparison. Moreover, because deep learning-based models had access to the same or more information about previous elements in the sequence at each prediction, neither recurrent neural networks nor transformer-based language models could achieve the desired level of improvement over classical N-gram models.    

In addition, obtained N-gram language models are superior to deep learning approaches regarding space complexity due to the fact that their model sizes are far smaller. When it comes to time complexity, the situation is again similar since deep learning methods require hyperparameter tuning, whereas N-gram language models require only a single training. When the results obtained are comprehensively taken into account, deep learning-based language models do not look more appropriate for the task than simple N-grams, and it seems like the data-hungry nature of deep learning models is still valid even for a case where we work on a language where the vocabulary size is minimal.

\section{Future Work}\label{sec6}
There is room for improvement in most scientific work. To further improve our work, some factors such as expecting developments from different areas of science, being able to access more powerful computational resources over time, and encouraging other people to work on similar problems by emphasizing possible outcomes of successful works on the studied problem ought to be remembered. 

Firstly, genomics is a relatively new discipline compared to other branches of science like mathematics. Because of this, it is very normal to expect that our understanding will get better over time, which means that conducting a similar study in the future could give far more significant insights. Also, a greater theoretical lower bound could be determined for the problem at hand over time. Coming up with such a reliable bound with our current knowledge was not easy for us.

Second of all, more detailed hyperparameter optimization strategies could give more intuition about why deep learning methods are not that successful. Of course, using powerful GPUs is essential for this, yet it could still be helpful for further studies.

Furthermore, expanding the dataset by using nucleotide sequences of genes existing in animals, plants and others could be an interesting work as well. Being able to find them might not be as simple as it was in this study and it requires huge computation resources such as GPUs, but the outcomes of the study could be far more general and conclusive.

Lastly, finding out more real-life problems strongly related to modeling nucleotide sequences of human genes could increase the scientific community's attention dramatically, which increases the amount of effort and investment put in to develop better strategies to model these incredible structures existing in nature, including our bodies.

\section*{Acknowledgments}
This work is partially supported by Boğaziçi University Research Fund under the Grant Number 16903 and by ERC grant (LifeLU, 101089287). Views and opinions expressed are however those of the author(s) only and do not necessarily reflect those of the European Union or the European Research Council Executive Agency. Neither the European Union nor the granting authority can be held responsible for them.

\bibliographystyle{unsrt}  
\bibliography{references}  

\newpage

\section*{Appendix}

\setcounter{table}{0}
\renewcommand{\tablename}{Table}
\renewcommand{\thetable}{S\arabic{table}}

\begin{table}[ht]
\centering
\caption{Hyperparameter experiments for rnn model}
\vskip\baselineskip
\begin{tabular}{|c|c|}
\hline
\textbf{Hyperparameters}                                                                                      & \textbf{Validation Perplexity} \\ \hline
\begin{tabular}[c]{@{}c@{}}Initial:\\ - EMBED\_DIM = 256\\ - LSTM\_DIM = 256\\ - NUM\_LAYERS = 2\end{tabular} & 3.5257                         \\ \hline
\begin{tabular}[c]{@{}c@{}}Change: \\ NUM\_LAYERS = 1\end{tabular}                                            & 3.5951                         \\ \hline
\begin{tabular}[c]{@{}c@{}}Change:\\ NUM\_LAYERS = 3\end{tabular}                                             & 3.4770                         \\ \hline
\begin{tabular}[c]{@{}c@{}}Change:\\ EMBED\_DIM = 128\end{tabular}                                            & 3.4995                         \\ \hline
\begin{tabular}[c]{@{}c@{}}Change:\\ LSTM\_DIM = 128\end{tabular}                                             & 3.6067                         \\ \hline
\begin{tabular}[c]{@{}c@{}}Change: \\ EMBED\_DIM = 512\end{tabular}                                           & 3.5441                         \\ \hline
\begin{tabular}[c]{@{}c@{}}Change:\\ LSTM\_DIM = 512\end{tabular}                                             & 3.4399                         \\ \hline
\begin{tabular}[c]{@{}c@{}}Change:\\ NUM\_LAYERS = 4\end{tabular}                                             & 3.4637                         \\ \hline
\begin{tabular}[c]{@{}c@{}}Change:\\ NUM\_LAYERS = 4 \\ EMBED\_DIM = 128 \\ LSTM\_DIM = 512\end{tabular}      & 3.3305                         \\ \hline
\begin{tabular}[c]{@{}c@{}}Change:\\ NUM\_LAYERS = 4 \\ LSTM\_DIM = 512\end{tabular}                          & \textbf{3.2726 (CHOSEN)}       \\ \hline
\begin{tabular}[c]{@{}c@{}}Change:\\ NUM\_LAYERS = 5 \\ LSTM\_DIM = 512\end{tabular}                          & 3.3610                         \\ \hline
\begin{tabular}[c]{@{}c@{}}Change:\\ NUM\_LAYERS = 4 \\ LSTM\_DIM = 1024\end{tabular}                         & 3.3656                         \\ \hline
\end{tabular}
\label{table:rnnhyp}
\end{table}

\newpage

\begin{table}[ht]
\centering
\caption{Hyperparameter experiments for transformer model}
\vskip\baselineskip
\begin{tabular}{|c|c|}
\hline
\textbf{Hyperparameters}                                                                                                                    & \textbf{Validation Perplexity} \\ \hline
\begin{tabular}[c]{@{}c@{}}Initial:\\ - EMBED\_DIM = 256\\ - FEED\_FORWARD\_DIM = 256\\ - NUM\_LAYERS = 2  \\ - NUM\_HEADS = 3\end{tabular} & 3.7018                         \\ \hline
\begin{tabular}[c]{@{}c@{}}Change: \\ NUM\_LAYERS = 3\end{tabular}                                                                          & 3.6932                         \\ \hline
\begin{tabular}[c]{@{}c@{}}Change:\\ NUM\_HEADS = 4\end{tabular}                                                                            & 3.7011                         \\ \hline
\begin{tabular}[c]{@{}c@{}}Change:\\ EMBED\_DIM = 512\end{tabular}                                                                          & 3.7189                         \\ \hline
\begin{tabular}[c]{@{}c@{}}Change:\\ FEED\_FORWARD\_DIM = 512\end{tabular}                                                                  & 3.7064                         \\ \hline
\begin{tabular}[c]{@{}c@{}}Change: \\ NUM\_LAYERS = 4\end{tabular}                                                                          & \textbf{3.6880 (CHOSEN)}       \\ \hline
\begin{tabular}[c]{@{}c@{}}Change:\\ NUM\_LAYERS = 5\end{tabular}                                                                           & 3.6903                         \\ \hline
\begin{tabular}[c]{@{}c@{}}Change:\\ EMBED\_DIM = 128\end{tabular}                                                                          & 3.7022                         \\ \hline
\begin{tabular}[c]{@{}c@{}}Change:\\ FEED\_FORWARD\_DIM = 128\end{tabular}                                                                  & 3.7095                         \\ \hline
\begin{tabular}[c]{@{}c@{}}Change:\\ NUM\_LAYERS = 4 \\ NUM\_HEADS = 4\end{tabular}                                                         & 3.6959                         \\ \hline
\end{tabular}
\label{table:trahyp}
\end{table}

\end{document}